\documentclass[a4paper,11pt]{article}
\usepackage{latexsym}
\usepackage{epsfig, subfigure, subfloat, graphicx, float }
\usepackage{amssymb}
\usepackage{amsmath}
\usepackage{epstopdf}
\usepackage{authblk}
\title{Onset of acceleration in a universe initially filled by dark and baryonic matters in nonminimally coupled teleparallel model }

\author[1]{H. Mohseni Sadjadi\thanks{mohsenisad@ut.ac.ir}}

\affil[1]{Department of Physics, University of Tehran}

\begin{document}

\maketitle

\begin{abstract}
 A non-minimally coupled quintessence dark energy in a teleparallel model of gravity is considered. It is clarified how a matter dominated universe with initial negligible dark energy density can evolve to a late time de Sitter space-time via the $Z_2$ symmetry breaking.
\end{abstract}

\section{Introduction}

Various models have been proposed to describe the present acceleration of the Universe, which is confirmed by analyzing different astrophysical data \cite{acc}.
 These models can be roughly classified into two classes:  i) Considering extra terms in the Einstein equations through the cosmological constant or exotic matter with negative pressure such as the quintessence scalar field.  ii) Modifying the Einstein theory of gravity such that at large scale the required negative pressure is provided. The “teleparallel” model of general relativity is one of these models in which one uses the curvatureless Weitzenbock connection instead of the usual Levi-Civita torsionless
connections \cite{tel}. In this context vierbeins are considered dynamical fields. The study of Universe acceleration in this framework has been the subject of many papers in recent years \cite{teleac}. There is some similarity between this model and ELKO \cite{elko} dark
energy model which may be related to the relation between the torsion and spinors.
\cite{Fabri}.

Another problem concerning dark energy models is the coincidence problem which states why nowadays the order of dark matter and dark energy densities, despite their different evolutions  are the same. This may be restated as why the dark energy density was so small and negligible at the earlier epochs.

In this paper we consider a spatially flat Friedmann-Lemaitre-Robertson-Walker (FLRW) Universe, and try to study the onset of acceleration in the framework of teleparallel gravity via the spontaneous symmetry breaking. We also take a look at the coincidence problem. To this end, we establish our model such that the scalar field is initially settled down at the minimum of its potential which we take to be zero. Indeed, the scalar field potential is assumed to have a $Z_2$ symmetry with a zero minimum value. An interaction source between dark matter and dark energy is required to hold the scalar field at this minimum.  As the nature of dark matter and dark energy has not yet been clarified (and as it is customary in the literature),  we consider the interaction only between dark sectors and do not consider nongravitational interaction between baryonic matter and dark sectors and so evade local gravitational tests.

In this way dark energy does not contribute in the total energy density of the Universe at earlier eras. But, if we want to relate onset of acceleration to symmetry breaking,  this procedure produces a problem in the minimal case: In the minimal case where the Friedmann equations have the same form as the usual general relativity, the acceleration cannot
occur in a scalar field model with negative potential \cite{Fara}. So if initially the potential is zero, as the symmetry breaking reduces the value of the potential, the acceleration will not happen in this model \cite{sad}. This was our motivation to consider a coupling between dark energy and scalar torsion and construct a non minimally coupled model.

Although teleparallel model of gravity with a non-minimally coupled scalar field to the scalar torsion has been mostly used in FLRW cosmology to study the present acceleration of the Universe, but it was also the subject of study in other situations: The post-Newtonian limit of this model was discussed in \cite{Li} and it was shown that the model is compatible with Solar system observations. As it was shown in \cite{Wei}, this feature is quite different from the usual scalar-tensor theories which require strict conditions on the parameter space to become consistent with the Solar System tests.
Spherically symmetric structures  supported by this model has also been the subject of some studies:  In \cite{Hor}, non-minimally coupled scalar field in the teleparallel framework was used to introduce a new class of boson stars with some novel characteristics. Asymptotically AdS hairy black hole solutions in three dimensions and in the context of teleparallel gravity were studied in \cite{Gonz}. Spherically symmetric solutions for models in which the scalar field is non-minimally coupled to the scalar torsion via a derivative coupling, was studied in \cite{Kofi}, where a new class of wormhole-like solutions was presented.

The plan of the paper is as follows:
In the second section we introduce the teleparallel model with two couplings: the coupling of the scalar field to the torsion scalar and the coupling of dark matter to dark energy. For the latter, beside the phenomenological method, we present a possible framework based on an action. In the third section we explain how these couplings hold the scalar field at the minimum of the effective potential at early and late times. Finally, we discuss the stability of the model in the fourth section.

 We use units $\hbar=c=8\pi G=1$.
\section{Coupled dark energy in the teleparallel model}

We consider a Universe filled with dark energy scalar field, $\phi$,  and pressureless dark and baryonic matters in the context of teleparallel model of gravity.
This model is described by the action \cite{sad3}
\begin{equation}\label{1}
S=\int ed^4x\left({T\over 2}+{1\over 2}(\partial_\mu \phi \partial^\mu \phi+\epsilon T\phi^2)-V(\phi)\right)+S_m.
\end{equation}
 $S_m$ indicates the contribution of baryonic and cold dark matters. In terms of the vierbeins, the metric tensor may be expressed as $g_{\mu \nu}=\eta_{BA} e_\mu^B e_\nu^A$, where  $\eta_{BA}=diag(1,-1,-1,-1)$.  $e$ is defined by $e=det(e_\nu^A)= \sqrt{-g}$ and $T$ is the torsion scalar
\begin{equation}\label{2}
T={1\over 2}T^{\alpha \beta \rho}T_{\rho \beta \alpha}+{1\over 4}T^{\alpha \beta \rho}T_{\alpha \beta \rho}-{T_{\alpha \beta}}^{\alpha} {T^{\alpha \beta}}_{\alpha},
\end{equation}
in which
${T^\alpha}_{\mu \nu}=e^{\alpha}_B(\partial_\mu e_\nu^B-\partial_\nu e_\mu^B)$.

Variation of (\ref{1}) with respect to $e^B_{\mu}$ gives the equations of motion
\begin{eqnarray}\label{3}
&&2(1+\epsilon\phi^2)\left({-{1\over 4}}e e_B^\mu T-ee^\rho_BT^\lambda_{\beta \rho}{S_\lambda}^{\mu \beta}+\partial_\nu(ee^\lambda_B{S_\lambda}^{\nu \mu})\right) \nonumber  \\
&-&ee^\mu_B\left({1\over 2}\partial_\nu\phi\partial^\nu \phi-V(\phi)\right)+ee^\nu_B\partial_\nu\phi\partial^\mu\phi+4\epsilon ee^\lambda_B{S_\lambda}^{\nu \mu}\phi\partial_\nu \phi\nonumber \\
&=&ee^\lambda_B{T^{(m)}}_\lambda^\mu.
\end{eqnarray}
${S_\rho}^{\mu \nu}$ is defined through
\begin{equation}\label{4}
{S_\rho}^{\mu \nu}={1\over 2}\left( \delta_\rho^\mu {T^{\lambda \nu}}_\lambda-\delta^\nu_\rho {T^{\lambda \mu}}_\lambda-{1\over 2}{T^{\mu \nu}}_\rho+{1\over 2}{T^{\nu \mu}}_\rho
+{1\over 2}{T_{\rho}}^{\mu \nu}\right),
\end{equation}
and ${T^{(m)}}_\lambda^\mu$ is the energy-momentum tensor corresponding to $S_m$ (not to be confused with the total energy momentum tensor):
\begin{equation}\label{r1}
e^{\lambda}_{j}{T^{(m)}}_\lambda^\mu={1\over e}{\delta(e\mathcal{L}^{(m)})\over \delta e^j_\mu},
\end{equation}
where $\mathcal{L}^{(m)}$ is the Lagrangian density associated to pressureless baryonic and dark matters.

To study the cosmological aspects of the model, we consider a spatially flat FLRW space-time
\begin{equation}\label{5}
ds^2=dt^2-a^2(t)(dx^2+dy^2+dz^2).
\end{equation}
for which $e^B_\nu=diag(1,a(t),a(t),a(t))$. $a(t)$ is the scale factor. The Hubble parameter is $H={\dot{a(t)}\over a(t)}$. For a homogeneous scalar field $\phi=\phi(t)$, from (\ref{3}), we obtain the Friedmann equations
\begin{eqnarray}\label{6}
3\left(1+\epsilon \phi^2\right)H^2&=&{1\over 2}\dot{\phi}^2+V(\phi)+\rho_{bm}+\rho_{dm}\nonumber \\
2\left(1+\epsilon \phi^2\right)\dot{H}&=&-\left(\dot{\phi}^2+4\epsilon H \phi \dot{\phi}+\rho_{bm}+\rho_{dm}\right),
\end{eqnarray}
where baryonic and cold dark matter energy densities are introduced by $\rho_{bm}$ and $\rho_{dm}$ respectively.

By defining effective energy density and pressure for the scalar field dark energy as
\begin{eqnarray}\label{7}
\rho_\phi&=&{1\over 2}\dot{\phi}^2+V(\phi)-3\epsilon H^2\phi^2\nonumber \\
P_\phi&=&{1\over 2}\dot{\phi}^2-V(\phi)+4\epsilon H \phi \dot{\phi}+\epsilon(3H^2+2\dot{H})\phi^2,
\end{eqnarray}
the Friedmann equations take their usual forms
\begin{eqnarray}\label{8}
H^2&=&{1\over 3}\left(\rho_\phi+\rho_{bm}+\rho_{dm}\right)\nonumber \\
\dot{H}&=&-{1\over 2}\left(\rho_\phi+P_\phi+\rho_{bm}+\rho_{dm}\right).
\end{eqnarray}
Variation of the action with respect to the scalar field gives
\begin{equation}\label{r2}
\ddot{\phi}+3H\dot{\phi}-\epsilon T\phi+V_{,\phi}=0.
\end{equation}
Therefore, from (\ref{r2}) one can see that the scalar field satisfies the continuity equation
\begin{equation}\label{r3}
\dot{\rho_\phi}+3H(\rho_\phi+P_\phi)=0.
\end{equation}
As $\rho_{bm}$ and $\rho_{dm}$ satisfy their own continuity equations, the total energy density, $\rho_T=\rho_\phi+\rho_{bm}+\rho_{dm}$, satisfies
\begin{equation}\label{r4}
\dot{\rho}_T+3H(\rho_T+P_\phi)=0.
\end{equation}

As the nature of dark sectors has not yet been clarified, it is legitimate to consider (possible) interaction between them. This idea was first introduced to alleviate the coincidence problem, that is to explain  why the order of magnitude of dark energy density is the same as that of dark matter, despite their different evolutions. In the minimal models in Einstein's theory of gravity this procedure has been employed in the literature to transmit energy density from the scalar field to dark matter, but in our case as we will show, this interaction serves to hold the scalar field at the zero level of its energy in earlier eras.

Following \cite{amendola} we consider an interaction between dark sectors, i. e. between the scalar field and cold dark matter, via the source $Q=f(\phi)T^{(dm)}\phi_{;\mu}$, where $T^{(dm)}$ is the trace of the energy momentum tensor of dark matter. Taking such an interaction (as we reveal) also has roots in Brans-Dicke models . General covariance implies that \cite{amendola}
\begin{eqnarray}\label{r5}
{T^{(dm)}}^\mu_{\nu;\mu}&=&f(\phi)T^{(dm)}\phi_{;\nu}\nonumber \\
{T^{(\phi)}}^\mu_{\nu;\mu}&=&-f(\phi)T^{(dm)}\phi_{;\nu}\nonumber \\
{T^{(bm)}}^\mu_{\nu;\mu}&=&0,
\end{eqnarray}
where ${T^{(dm)}}^\mu_{\nu}, {T^{(\phi)}}^\mu_{\nu},  {T^{(bm)}}^\mu_{\nu}$ are the energy momentum tensor of dark matter, quintessence, and baroynic matter respectively. For homogenous scalar field $\phi=\phi(t)$, (\ref{r5}) gives
\begin{equation}\label{9}
\dot{\rho}_\phi+3H(P_\phi+\rho_\phi)=-\dot{\phi}f(\phi)\rho_{dm},
\end{equation}
and
\begin{equation}\label{10}
\dot{\rho}_{dm}+3H\rho_{dm}=\dot{\phi}f(\phi)\rho_{dm}.
\end{equation}
{\it Note that we have not considered any non-gravitational interaction between dark sectors and baryonic matter, so relax any local gravitational constraints}, therefore  $\rho_{bm}$ satisfies as usual
\begin{equation}\label{11}
\dot{\rho}_{bm}+3H\rho_{bm}=0,
\end{equation}
whose solution is given by $\rho_{bm}=\rho_{bm0}a^{-3}$,  where we take the present scale factor as $a_0=1$. Subscript $0$ denotes the value of a parameter at the present time.

It is interesting to note that our chosen interaction may also be obtained from an action.  To see this, we consider the action
\begin{equation}\label{15}
S=\int ed^4x\left({T\over 2}+{1\over 2}(\partial_\mu \phi \partial^\mu \phi+\epsilon T\phi^2)-V(\phi)\right)+S_m[\tilde{{e}_\mu^B}^{(i)}, \psi^{(i)}],
\end{equation}
where $\psi^{(i)}$ is the $ith$ matter component that is baryonic or dark matter, and $\tilde{{e}_\mu^B}^{(i)}$ is specified by
the conformal transformation
\begin{equation}\label{c}
{\tilde{{e}_\mu^B}^{(i)}=A^{(i)}(\phi)e^B_\mu,\,\,    \tilde{{e}^\mu_B}}^{(i)}={\left(A^{(i)}(\phi)\right)}^{-1}e^\mu_B,\,\,  \tilde{e}^{(i)}=\left(A^{(i)}(\phi)\right)^4e.
\end{equation}
Considering different conformal coupling for different species violates the weak equivalence principle but (in the presence of not very well known components) is customary in the literature, e.g. in the mass varying neutrino model \cite{neutrino}, or in the model of variable gravity \cite{variable}. For ordinary baryonic matter we take $A^{(i=bm)}(\phi)=1$ and for cold dark matter $A^{(i=dm)}(\phi)=A(\phi)$. The transformed metric in the {\it dark matter} sector is then given by $\tilde{g}_{\mu \nu}=A^2(\phi)g_{\mu \nu}$.
Using \cite{Dong}
\begin{equation}
{\delta (\tilde{e} \mathcal{L}(\tilde{e} \psi))\over \delta \tilde{e}^B_\alpha}=\tilde {e}\tilde{e}^\rho_B{\tilde{T}^\alpha}_\rho
\end{equation}
 where $\tilde{T}^\alpha_\rho$ is the energy momentum tensor in the tilde frame, variation of the action with respect to $\phi$ gives
\begin{equation}\label{16}
\Box^2\phi+V_{,\phi}-\epsilon T \phi+ A^3(\phi)A_{,\phi}\tilde{{T}^{\mu}_{\mu}}^{(dm)}=0,
\end{equation}
which for pressureless dark matter becomes
\begin{equation}\label{r6}
\Box^2\phi+V_{,\phi}-\epsilon T \phi+ A^{-1}A_{,\phi}\rho_{dm}=0.
\end{equation}
To find this we used $\tilde{T}^{(dm)}={T^{(dm)}\over A^4}$ giving $A^4\tilde{\rho}_{dm}=\rho_{dm}$. Equation (\ref{r6}) is the same as (\ref{9}) provided that we take \cite{sad2}
\begin{equation}\label{r7}
A(\phi)=\exp\left(\int f(\phi)d\phi\right).
\end{equation}

Variation of the action with respect to the tetrad $e^B_\mu$ gives
\begin{eqnarray}\label{17}
&&2(1+\epsilon\phi^2)\left({-{1\over 4}}e e_B^\mu T-ee^\rho_BT^\lambda_{\beta \rho}{S_\lambda}^{\mu \beta}+\partial_\nu(ee^\lambda_B{S_\lambda}^{\nu \mu})\right) \nonumber  \\
&-&ee^\mu_B\left({1\over 2}\partial_\nu\phi\partial^\nu \phi-V(\phi)\right)+ee^\nu_B\partial_\nu\phi\partial^\mu\phi+4\epsilon ee^\lambda_B{S_\lambda}^{\nu \mu}\phi\partial_\nu \phi\nonumber \\
&=&ee^\lambda_B{T^{(bm)}}_\lambda^\mu+ A^4(\phi)ee^{\rho}_B \tilde {{T}^\nu_\rho}^{(dm)}
\end{eqnarray}
which, by substituting $\rho_{dm}=A^4(\phi)\tilde{\rho}_{dm}$, yields our previous Friedmann equations.
Therefore the equations of motion may be rederived via an action context. Both approaches are phenomenological (resulting interactions have not base in a fundamental theory) although the second approach has roots in Brans-Decke theory of gravity \cite{amendola},\cite{kh}.

Hereinafter, we use a rescaled dark matter energy density, which is rather a mathematical object, defined by
\begin{equation}
\rho_=A(\phi)\rho_{dm},
\end{equation}
to simplify (\ref{r6}) and (\ref{10}) as
\begin{eqnarray}\label{13}
&&\ddot{\phi}+3H\dot{\phi}+6\epsilon H^2\phi+V_{,\phi}+A_{,\phi}\rho=0\nonumber\\
&&\dot{\rho}+3H\rho=0.
\end{eqnarray}
To derive (\ref{13}) we used also $T=-6H^2$ which holds in FLRW space-time. So $\phi$ evolves in an effective potential specified by
\begin{equation}\label{14}
V^{eff.}_{,\phi}=6\epsilon H^2\phi+V_{,\phi}+A_{,\phi}\rho,
\end{equation}
where $\rho$ is simply given by $\rho=\rho_0a^{-3}$. This simplicity will help us establish our model.

Note that the interaction between dark sectors does not enter in the Friedmann equations explicitly, but modifies the continuity equations of dark components.
In the following we use (\ref{13}) and the Friedmann equations (\ref{6}) to study the positive acceleration of the Universe.

\section{Deceleration to acceleration phase transition via $Z_2$ symmetry breaking}

The second derivative of the scale factor, i. e. $\ddot{a}$, has the same sign as
\begin{equation}\label{18}
-qH^2=\dot{H}+H^2={-2\dot{\phi}^2+2V-\rho_{bm}-A\rho-12\epsilon H\phi \dot{\phi}\over 6(1+\epsilon \phi^2)},
\end{equation}
where $q$ is the deceleration parameter. So for $\dot{H}+H^2 >(<)0$ we have an accelerated (decelerated) Universe. We aim to establish a model where initially the scalar field was settled down at the zero minimum of its potential such that it had no contribution in the energy density of the Universe.
This is motivated by the coincidence problem which states why the dark matter and dark energy, despite their different evolutions have the same order today or in other words why the dark energy density was much smaller than dark matter at early times. Besides, we require that the acceleration becomes positive in the present epoch. For our goal we try to use the symmetry breaking process. To do so we choose the well-known $Z_2$ symmetric potential
\begin{equation}\label{19}
V(\phi)=-{\mu^2\over 2}\phi^2+{\lambda\over 4}\phi^4,
\end{equation}
and consider \cite{kh}
\begin{equation}\label{20}
A(\phi)=1+{\gamma\over 2}\phi^2+\mathcal{O}\left(\gamma ^2 \phi^4\right).
\end{equation}
$\lambda$ and $\mu$ are positive constants.  Evenness of $A(\phi)$  guarantees $Z_2$ symmetry of the effective action. $\gamma>0$ is assumed as an inverse mass squared scale such that $\phi^2\ll {1\over \gamma}$. Therefore terms containing higher power of $\phi$ are absent in the polynomial supposed in(\ref{20}). The chosen form for $A(\phi)$ allows the symmetry breaking via the scale factor evolution.

To see how this model can express an initial matter dominated decelerated Universe that precedes an acceleration phase arisen from the symmetry breaking, let us write the equation of motion of the scalar field as
\begin{equation}\label{21}
\ddot{\phi}+3H\dot{\phi}+V^{eff.}_{,\phi}=0,
\end{equation}
where the effective potential is
\begin{equation}\label{22}
V^{eff.}_{,\phi}=(6\epsilon H^2-\mu^2+\gamma \rho)\phi+\lambda \phi^3.
\end{equation}
So we can define an effective mass  by
\begin{equation}\label{23}
\mu_{eff.}^2=\gamma \rho+6\epsilon H^2-\mu^2.
\end{equation}
We assume that initially the field was settled down at the  point $\phi=0$ which is the solution of the field equation and is the minimum of the effective potential. In this era
$A(\phi)=1$ and $\rho$ defined by (\ref{r7}) equals $\rho_{dm}$, hence $H^2={\rho+\rho_{bm}\over 3}$ and
\begin{equation}\label{24}
\mu_{eff.}^2=\left(\gamma \rho_{0}+2(\rho_{0}+\rho_{bm0})\epsilon\right)a^{-3}-\mu^2.
\end{equation}
If
\begin{equation}\label{r8}
\left(\gamma \rho_{0}+2(\rho_{0}+\rho_{bm0})\epsilon\right)>0,
\end{equation}
then when matter densities are large enough ($a$ is small enough), the squared of the effective mass term is positive and the shape of the effective potential is concave and $\phi=0$ is a stable point. To satisfy (\ref{r8}) for $\epsilon<0$, we must have $\gamma> 0$. Based on these remarks, we can model our theory such that the field was settled down initially at the  point $\phi=0$ which is the solution of (\ref{13}) and also is the minimum of the effective potential. In this era the Universe was matter dominated and was in deceleration phase
\begin{equation}\label{25}
\dot{H}+H^2={-\rho_{bm}-\rho\over 6}<0.
\end{equation}
So initially the dark energy did not contribute in the Universe's ingredients: $\Omega_\phi=0$. The relative densities are defined through $\Omega_i={\rho_i\over 3H}$.
In this era, the ratio of matter energy densities was a constant
\begin{eqnarray}\label{26}
{\Omega_{bm}\over \Omega_{dm}}&=&{\rho_{bm}\over \rho_{dm}}\nonumber \\
&=&{\rho_{bm0}\over \rho_{0}},
\end{eqnarray}
To derive the second equality, we used $\rho_{dm}=\rho$ which holds when $\phi=0$. During its evolution the scale factor increases and becomes greater than a critical value $a>a_c$  specified by
\begin{equation}\label{27}
a_c=\left({\gamma \Omega_{0}+2(\Omega_{0}+\Omega_{bm0})\epsilon \over {\mu^2\over 3H_0^2}}\right)^{1\over 3},
\end{equation}
where $\Omega_0={\rho_0\over 3H_0}$ and $\Omega_{bm0}={\rho_{bm0}\over 3H_0}$, and subscript $0$ as before denotes the value of a parameter at the present era. When this happens, the squared of the effective mass becomes negative and the tachyonic scalar field leaves the previous vacuum, i. e.   $\phi=0$ (which is now the new local maximum of the effective potential), and rolls down towards the new stable vacuum of the system with a rate of order $\mu_{eff.}$ and the $Z_2$ symmetry breaks.

It is worth noting that in the minimal model $\epsilon=0$, after the symmetry breaking the scalar field rolls down its potential and $V$ becomes negative forbidding the Universe to accelerate (see (\ref{18})). {\it{Therefore $\epsilon\neq 0$ is necessary for onset of acceleration}}.

The minimum of the effective potential is derived from
\begin{equation}\label{29}
V^{eff.}_{,\phi}=(6\epsilon H^2-\mu^2+\gamma \rho_{dm})\phi+\lambda \phi^3=0.
\end{equation}
 When the scale factor crosses the limit (\ref{27}), the scalar field rolls towards the minimum of the effective action and  begins a damped oscillation around it and finally tends to $\phi_c$ specified by (\ref{29}). Using (\ref{11}) and (\ref{13}) we deduce that  $\rho_{bm}$ and $\rho$ are decreasing functions of time. Hence, we expect that if $H$ tends to a non-zero constant, they approach zero at late time. Whence we expect the following consistent solution for the system at late time
\begin{eqnarray}\label{30}
&&(6\epsilon H_c^2-\mu^2)\phi_c+\lambda \phi_c^3=0\nonumber \\
&&\rho_{bm}=0,\,\,\, \rho_{dm}=0\nonumber \\
&&H_c^2={-{\mu^2\over 2}\phi_c^2+{\lambda \over 4}\phi_c^4\over 3(1+\epsilon \phi_c^2)},
\end{eqnarray}
 which specifies a de Sitter space time. For the minimal case $\epsilon=0$, or the Einstein theory of gravity we have $\phi_c^2={\mu^2\over \lambda}$ which yields $H_c^2= -{\mu^4\over 4\lambda}$. Thus solution (\ref{30}) is not acceptable in the minimal case.  For $\epsilon>0$ either $\phi^2_c$ or $H_c^2$ is negative so we restrict ourselves to $\epsilon<0$. This reveals the importance of $\gamma\neq 0$ for initial vanishing of dark energy (see the discussion after (\ref{r8})).

The solution of (\ref{30}) is given by
\begin{equation}\label{31}
\phi_c^2={2\epsilon\mu^2-\lambda\pm \sqrt{4\epsilon^2\mu^4+2\epsilon \lambda \mu^2+\lambda^2}\over 3\epsilon \lambda}.
\end{equation}
So after leaving the first vacuum $\phi=0$, the scalar field goes towards the new minimum of the effective potential, begins a damped oscillation around it and eventually settles down at $\phi_c$.

\section{Dynamical analysis}
 To investigate the evolution of the system more precisely late time and to examine the stability condition for the fixed points we perform a dynamical phase analysis. By defining the variables
$x={\dot{\phi}\over \sqrt{6}H},\,\, y=\phi,\,\, u={\sqrt{\rho}\over \sqrt{3}H},\,\, v={\sqrt{\rho_{bm}}\over \sqrt{3} H}$,
we find an autonomous system of differential equations
\begin{eqnarray}\label{32}
&&x'=-3x-\sqrt{3\over 2}\left(1-x^2-v^2-A(y)u^2+\epsilon y^2\right)f(y)-\sqrt{3\over 2}A_{,y}u^2 \nonumber \\
&&-\sqrt{6}\epsilon y-sx=:E(x,y,u,v)\nonumber \\
&&y'=\sqrt{6}x \nonumber \\
&&u'=:F(x,y,u,v)=-{3\over 2}u-su \nonumber \\
&&v'=:G(x,y,u,v)=-{3\over 2}v-sv
\end{eqnarray}
where $'$ denotes a derivative with respect to $\ln(a)$,
\begin{equation}
s={{\dot H}\over H^2}=-{3\over 2}{6x^2+4\sqrt{6}\epsilon yx+3A(y)u^2+3v^2\over 3\epsilon y^2+3},
\end{equation}
and $f(y)={V_{,\phi}\over V}$. Fixed points are obtained as
\begin{equation}\label{33}
I: \{ \epsilon=-{f(\bar{y})\over \bar{y}(\bar{y}f(\bar{y})+2)},\bar{u}=0,\bar{v}=0,\bar{x}=0 \},
\end{equation}
\begin{equation}\label{34}
II: \{ \epsilon=-{A_{,y}(\bar{y})\over \bar{y}(\bar{y}A(\bar{y})+2A(\bar{y}))},\bar{u}={2\over \sqrt{2\bar{y}A_{,y}(\bar{y})+4A(\bar{y})} },\bar{v}=0,\bar{x}=0\},
\end{equation}
\begin{equation}\label{35}
III: \{\epsilon=-{A_{,y}(\bar{y})\bar{u}^2\over \bar{y}}, \bar{v}={1\over 2}\sqrt{4-2A_{,y}(\bar{y})\bar{u}^2\bar{y}-4A(\bar{y})\bar{u}^2},\bar{x}=0\},
\end{equation}
\begin{equation}\label{36}
IV: \{\bar{u}=0, \bar{v}=1,\bar{x}=0, \bar{y}=0\}.
\end{equation}
 There exists also an additional condition coming from the Friedmann equation
 \begin{equation}\label{37}
 z-\epsilon y^2+A(y)u^2+v^2+x^2=1,
 \end{equation}
 where $z={V(y)\over 3H^2}$. The deceleration parameter at the fixed points is
 \begin{equation}\label{38}
 q=\left({1\over 1+\epsilon \bar{y}^2}\right)\left(\bar{z}+{\bar{v}\over 2}+{A(\bar{y})\bar{u}\over 2}\right).
 \end{equation}

  After some computations,  we find that at point II, $\epsilon=-{\gamma\over 2(1+\gamma \bar{y}^2)}$ and $\bar{u}={1\over \sqrt{\gamma \bar{y}^2+1}}$.
 So we deduce $1+\epsilon \bar{y}^2>0$ and (\ref{37}) leads to  $\bar{z}=0$ resulting in a deceleration phase. Similarly for point III, we obtain
 $\epsilon=-{1\over 2}\gamma \bar{u}^2$ and $\bar{v}^2=1-\bar{u}^2(1+\gamma \bar{y}^2)$ resulting in $1+\epsilon \bar{y}^2>0$. $\bar{z}=0$  holds in this case too. So accompanied by IV, points II and III can be classified in a group describing a decelerated Universe, $q>0$,  dominated by baryonic and dark matter.

Now consider point I.   By equating $\phi_c$ with $\bar{y}$, we can derive $\epsilon=-{f(\bar{y})\over \bar{y}(\bar{y}f(\bar{y})+2)}$ from (\ref{30}). Hence point $I$ is the same as the critical point considered in the previous section. The stability of this fixed point can be examined by studying small perturbations around the fixed points:  $x=\bar{x}+\delta x,\,\, y=\bar{y}+\delta y,\,\, u=\bar{u}+\delta u,\,\,\, v=\bar{v}+\delta v$. We obtain
\begin{equation}\label{39}
{d\over d{\ln a}}{ \left( \begin{array}{cccc}
\delta x\\
\delta y \\
\delta u \\
\delta v
\end{array} \right)}=\mathcal{M} \left( \begin{array}{cccc}
\delta x\\
\delta y \\
\delta u \\
\delta v
\end{array} \right),
\end{equation}
where

\begin{equation}\label{40}
\mathcal{M}=\left( \begin {array}{cccc} E_{,x}& E_{,y} &E_{,u}&E_{,v}  \\ \noalign{\medskip} \sqrt{6} &0&0&0\\ \noalign{\medskip}F_{,x}&F_{,y}&F_{,u}&F_{,v}
\\ \noalign{\medskip}G_{,x}&G_{,y}&G_{,u}&G_{,v}\end {array} \right).
\end{equation}
Eigenvalues of $\mathcal{M}$ for point I are derived as
\begin{eqnarray}\label{41}
-{3\over 2},\,\,\, -{3\over 2},\,\,\, -{3\over 2}+{1\over 2}\sqrt{9-12 \epsilon y^2f_{,y}-24\epsilon y f-12f_{,y}-24\epsilon}, \nonumber \\
-{3\over 2}-{1\over 2}\sqrt{9-12 \epsilon y^2f_{,y}-24\epsilon y f-12f_{,y}-24\epsilon}
\end{eqnarray}
In the other points $\mathcal{M}$ has a positive eigenvalue  (which equals 3).  If the (real parts) of the eigenvalues (\ref{41}) are negative then point I is stable. This is satisfied for
\begin{equation}\label{42}
\epsilon y^2f_{,y}+2\epsilon y f+f_{,y}+2\epsilon>0.
\end{equation}

Substituting $\bar{y}$ from (\ref{31}) in the above inequality, we find that point I is stable for
\begin{eqnarray}\label{43}
&&\mp\sqrt {4\,{\epsilon}^{2}{\mu}^{4}+2\,\epsilon\,\lambda\,{\mu}^{2}+{
\lambda}^{2}} \left( 2\,{\epsilon}^{2}{\mu}^{4}-2\,\epsilon\,\lambda\,
{\mu}^{2}-{\lambda}^{2} \right)\nonumber \\
&& -4\,{\epsilon}^{3}{\mu}^{6}-6\,{\mu}^{
4}{\epsilon}^{2}\lambda-3\,\epsilon\,{\lambda}^{2}{\mu}^{2}-{\lambda}^
{3}>0.
\end{eqnarray}
Therefore, if we choose our parameters such that (\ref{43}) is satisfied, all eigenvalues in (\ref{41})become negative and (\ref{30}) describes an attractor solution, towards which the system tends. But we remember that among the initial situations, the coincidence problem may be alleviated when $\phi$ was settled down at $\phi=0$ initially (i.e.  before that the scalar factor crossed $a_c$ in (\ref{27})).

\begin{figure}[H]
\centering\epsfig{file=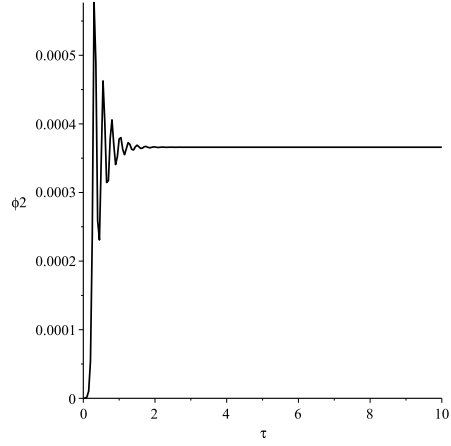,width=6cm,angle=0}
\caption{ An illustration of $\phi^2$  in terms of $\tau=tH_0$ for $\{{\lambda\over H_0^2} =10^6,{\mu\over H_0}=2, \epsilon=-10,\gamma=0.005\}$ and with the initial conditions
$\{\phi(0)=0,{d{\phi}\over d\tau}(0)=0.01,{\rho_{dm}(0)\over H_0^2}=7.5,{\rho_{bm}(0)\over H_0^2}=1.5\}$.} \label{fig1}
\end{figure}

To get more physical intuition about the model and its viability, let us depict some figures via numerical methods showing the evolution of the system .
The evolution of the quintessence is depicted in Fig. (\ref{fig1}) using the equations of motion ((\ref{11}), (\ref{13}))and the Friedmann equations. To depict this figure, we assumed that the velocity of the scalar field slightly deviates from zero due to fluctuation around the unstable point $\phi=0$ after the symmetry breaking.
Figure (\ref{fig1})shows that the scalar field leaves $\phi=0$ after the symmetry breaking and tends towards the critical point about which it begins a damped oscillation and finally settles down in a time of order $1\over H_0$.

\begin{figure}[H]
\centering\epsfig{file=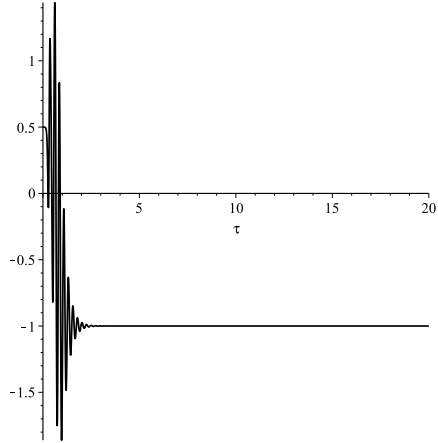,width=6cm,angle=0}
\caption{Deceleration parameter in terms of $\tau=tH_0$ for $\{{\lambda\over H_0^2} =10^6,{\mu\over H_0}=2, \epsilon=-10,\gamma=0.005\}$ and with the initial conditions
$\{\phi(0)=0,{d{\phi}\over d\tau}(0)=0.01,{\rho_{dm}(0)\over H_0^2}=7.5,{\rho_{bm}(0)\over H_0^2}=1.5\}$.} \label{fig2}
\end{figure}

\begin{figure}[H]
\centering\epsfig{file=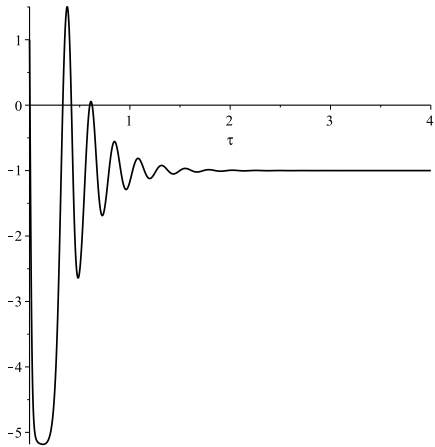,width=6cm,angle=0}
\caption{The equation of state parameter of the scalar field in terms of $\tau=tH_0$ for $\{{\lambda\over H_0^2} =10^6,{\mu\over H_0}=2, \epsilon=-10,\gamma=0.005\}$ and with the initial conditions
$\{\phi(0)=0,{d{\phi}\over d\tau}(0)=0.01,{\rho_{dm}(0)\over H_0^2}=7.5,{\rho_{bm}(0)\over H_0^2}=1.5\}$.} \label{fig3}
\end{figure}

In figure (\ref{fig2}) the deceleration parameter is depicted using the Friedmann equation and ((\ref{11}), (\ref{13})), showing that in a time of order the Hubble time the Universe experiences transitions from deceleration to the acceleration phase and remains finally in the acceleration phase.
The equation of state parameter of the scalar field is illustrated in fig(\ref{fig3}) showing that, at late lime, this parameter is near $w_\phi=-1$, which is consistent with observations \cite{planck}. Note that crossing the phantom divide line by the scalar field in the nonmimimal teleparallel model was discussed in the literature \cite{sad3}.

\section{Summary}
A spatially flat FLRW space-time was considered in the teleparallel model of gravity. We tried to describe the onset of late time acceleration via $Z_2$ symmetry breaking. We related this phenomenon to coupling of the quintessence to the torsion.

 Attributing the acceleration to $Z_2$ symmetry breaking was studied in the hybrid quintessence model as well as in the symmetron model in the usual Einstein model of gravity \cite{kh}. However, in these models as, via the symmetry breaking procedure, the scalar field rolls down its potential after its initial stay,  the potential decreases while the kinetic energy increases and this is not in favor of acceleration, this can be seen from relation (\ref{18})(with $\epsilon=0$ this relation has the same form as the Einstein model of gravity). In other words, in these models,  it would be better for acceleration that the field stay at its initial position (with a positive potential) instead of moving via the symmetry breaking. So, in these theories,  if even a positive acceleration happens (as it is confirmed numerically in \cite{kh}), it is driven by a positive term included initially in the potential playing the role of a cosmological constant and has nothing to do with the symmetry breaking which even reduces the potential\cite{sad}. This acceleration ends as the quintessence overshoots and oscillates about the new minimum of the effective potential.

In the coupled teleparallel model the situation changes extremely. As the deceleration factor takes a new form in the non minimally coupled teleparallel model (see (\ref{18})), a negative coupling provides the conditions required for acceleration, and in contrast to the minimal model drives the Universe to a de Sitter space time in the late time. So the deceleration to acceleration phase occurs. Such a transition to a de Sitter space-time is forbidden in the minimal case (see the discussion after (\ref{30})). The other feature of our model is that the acceleration begins in a Universe in which the dark energy density  had zero contribution initially. This is due to the coupling between dark energy and scalar torsion providing a concave effective potential and consequently providing a positive squared effective mass in the early eras (see \ref{23}).

Conditions to have attractor solutions were also derived in (\ref{43}). The evolution of the deceleration factor and the scalar field and also its equation of state parameter were depicted numerically showing the possibility of having a stable acceleration phase in a time of order of the present Hubble time.

\end{document}